\documentclass[11pt]{article}
\usepackage[utf8]{inputenc}
\usepackage{mathrsfs}  
\usepackage{titlesec}

\titleformat*{\section}{\large\bfseries}
\usepackage{setspace}
\usepackage{amssymb,amsmath,amsthm,amsfonts,amstext}
\usepackage[english]{babel}
\usepackage[left=1.5cm,right=1.5cm,top=1.5cm,bottom=1.5cm]{geometry}
\usepackage{enumerate}
\usepackage[mathscr]{euscript}

\usepackage{mathtools,braket}
\usepackage{bm, hyperref}
\usepackage{xcolor, float}
\usepackage[capitalize]{cleveref}

\newtheorem{theorem}{Theorem}
\newtheorem{corollary}[theorem]{Corollary}
\newtheorem{proposition}[theorem]{Proposition}
\newtheorem{lemma}[theorem]{Lemma}

\newtheorem{definition}{Definition}

\title{\Large \textbf{Algorithms for 2-connected network design and
	flexible Steiner trees with a constant number of terminals}
}

\author{\large
Ishan Bansal\thanks{
	{\tt ib332@cornell.edu}. Cornell University, Ithaca, NY, USA.}
\and
Joseph Cheriyan\thanks{
{\tt jcheriyan@uwaterloo.ca}.
	Department of Combinatorics and Optimization, University of Waterloo, Canada.}
\and
Logan Grout\thanks{
	{\tt lcg58@cornell.edu}.
	Cornell University, Ithaca, NY, USA.}
\and 
Sharat Ibrahimpur\thanks{
	{\tt sharat.ibrahimpur@uwaterloo.ca}.
	Department of Combinatorics and Optimization, University of Waterloo, Canada.}
}
\date{}

\usepackage{tikz}
\usepackage{hyperref}

\usepackage[mathscr]{euscript}

	\newcommand{\Rp}{\ensuremath{\mathbb R_{\geq 0}}}

	\newcommand{\fpt}{\ensuremath{\mathbf{FPT}}}

	\newcommand{\opt}{\textsc{opt}}
	\newcommand{\optsoln}{\textsc{optsoln}}
	\newcommand{\safe}{\mathscr{S}}
	\newcommand{\unsafe}{\mathscr{U}}

	\newcommand{\cp}{\ensuremath{\widehat{\cal P}}}

	\newcommand{\fst}{\mathrm{FST}}
	\newcommand{\term}{\ensuremath{T}}
	
	\newcommand{\pterm}{\ensuremath{\widetilde{\term}}}
	\newcommand{\kfst}[1][k]{#1\text{-}\fst}
	\newcommand{\twofst}[1][2]{#1\text{-}\fst}

	\newcommand{\intn}{\mathrm{int}}

	\newcommand{\kcycle}[1][k]{#1\text{-}\mathrm{Steiner}\text{-}\mathrm{cycle}}
	\newcommand{\scycle}{\mathrm{Steiner}\text{-}\mathrm{cycle}}
	\newcommand{\spath} {\mathrm{Steiner}\text{-}\mathrm{path}}
	\newcommand{\ktwoecs}[1][k]{#1\text{-}\mathrm{Steiner}\text{-}\mathrm{2ECS}}
	\newcommand{\ktwoncs}[1][k]{#1\text{-}\mathrm{Steiner}\text{-}\mathrm{2NCS}}

	\newcommand{\A}{\mathcal{A}}

	\newcommand{\tc}{\tilde{c}}
	\newcommand{\tG}{\widetilde{G}}
	
	\newcommand{\tV}{\widetilde{V}}
	\newcommand{\tQ}{\widetilde{Q}}
	\newcommand{\tE}{\widetilde{E}}

	\newcommand{\hF}{\hat{F}}
	\newcommand{\hH}{\hat{H}}

	\newcommand{\mG}[1]{{G_{#1}^{\min}}}
	\newcommand{\fptas}{FPTAS}

\newcommand{\Gamblocks}[1]{\Gamma^{\mathit{blocks}}_{#1}}

\newcommand{\bhtcycle}{\text{BHT-cycle}}
\newcommand{\bhtpath}{\text{BHT-path}}

\newcommand{\BT}[1]{B(#1)}
\newcommand{\CBT}[1]{\ensuremath{\hat{B}(#1)}}
\newcommand{\ahmspath}{\text{1-protected path}}
\newcommand{\ahmspaths}{\text{1-protected paths}}

\newcommand{\ve}{\varepsilon}

\begin{document}

\maketitle
%%%%%%%%%%%%%%%%%%%%%%%%%%%%%%%%%%%%%%%%%%%%%%%%%%%%%%

\begin{abstract}
{
The $k$-Steiner-2NCS problem is as follows:
Given a constant $k$, and an undirected connected graph $G = (V,E)$,
non-negative costs $c$ on the edges, and
a partition $(\term, V\setminus{\term})$ of $V$ into a set of terminals, $\term$, 
and a set of non-terminals (or, Steiner nodes),
where $|\term|=k$,
find a min-cost two-node connected subgraph that contains the terminals.

We present a randomized polynomial-time algorithm for the unweighted problem, 
and a randomized {\fptas} for the weighted problem.

We obtain similar results for the $k$-Steiner-2ECS problem, where the input 
is the same, and the algorithmic goal is to find a min-cost two-edge 
connected subgraph that contains the terminals.

Our methods build on results by Bj\"orklund, Husfeldt, and Taslaman (SODA~2012)
that give a randomized polynomial-time algorithm for the unweighted 
$\kcycle$ problem; this problem has the same inputs as the 
unweighted $k$-Steiner-2NCS problem, and the algorithmic goal is to find a min-cost 
simple cycle $C$ that contains the terminals
($C$ may contain any number of Steiner nodes).
}
\end{abstract}

\vfill
\newpage

%%%%%%%%%%%%%%%%%%%%
\section{Introduction} \label{sec:intro}
{ 
We address the following open question in the area of network design: Is 
there a polynomial-time algorithm for the $\ktwoncs$ problem (where the 
number of terminals, denoted by $k$, is a constant)? The input is an 
undirected connected graph $G = (V,E)$, non-negative costs $c$ on the 
edges, and a partition $(\term, V\setminus{\term})$ of $V$ into terminals ($\term$) 
and non-terminals (or, Steiner nodes). The algorithmic goal is to find 
a min-cost two-node connected subgraph that contains the terminals.
See Feldman et~al.~\cite{FML22}, the third item in the last section (open problems),
and for related results, see \cite{Jordan03,FR06}.

We present a randomized polynomial-time algorithm for the unweighted $\ktwoncs$ problem, 
and a randomized {\fptas} for the weighted problem.

We obtain similar results for the $\ktwoecs$ problem, where the input 
is the same, and the algorithmic goal is to find a min-cost two-edge 
connected subgraph that contains the terminals,
see Section~\ref{sec:fptas-k-fst}.
In fact we address a generalization of the $\ktwoecs$ problem called
the \textit{k-Flexible Steiner Tree} problem ($\kfst$):
The input consists of an undirected graph $G = (V,E)$ with nonnegative costs
on the edges $c\in\Rp^E$, a partition of the edge-set $E$ into a
set $\safe$ of \textit{safe edges} and a set $\unsafe$ of \textit{unsafe
edges}, and a subset of the nodes $\term \subseteq V$ with $|\term|=k$,
called the \textit{terminals}.  The algorithmic goal is to find a
connected subgraph $H=(U,F)$ of minimum cost (i.e., minimizing
$c(F)$) such that $T\subseteq{U}$ and for any unsafe edge $e\in{F}$,
the graph $H-e$ is connected.
Clearly, the $\ktwoecs$ problem is a special case of $\kfst$ where all edges are unsafe.

Our methods build on results by Bj\"orklund, Husfeldt, and Taslaman (SODA~2012) \cite{BHT12}
that give a randomized polynomial-time algorithm for the unweighted 
$\kcycle$ problem; this problem has the same inputs as the 
$\ktwoncs$ problem, and the algorithmic goal is to find a min-cost 
simple cycle $C$ that contains the terminals
($C$ may contain any number of Steiner nodes).
To the best of our knowledge, no polynomial-time (deterministic or randomized) algorithm is
known for finding an optimal solution of the weighted $\kcycle$ problem, even for $k=3$;
this problem has been open for several decades, see \cite{Dean91}.
There are a number of related results from the last few decades; for example, see
\cite{FW92,CLNV14}.

The randomized polynomial-time algorithm for the unweighted $\kcycle$ problem of \cite{BHT12}
extends easily to a randomized FPTAS for the weighted $\kcycle$ problem,
by using techniques from Ibarra \& Kim \cite{IK:jacm75} and
Hochbaum \& Shmoys \cite{HS:jacm86}, see Proposition~\ref{propos:BHTweighted}.
Our randomized FPTAS results (discussed above) use the same methods,
see Corollaries~\ref{coro:2ncs-weighted} and~\ref{coro:k-fst-weighted}.
%
%	We use the $\tilde O()$ notation to hide poly-logarithmic factors
%	when estimating the running time of algorithms, i.e., $\tilde O(f(n))
%	= O(f(n) \log^j{n})$ for a positive integer $j=O(1)$.
}
%%%%%%%%%%%%%%%%%%%%
\section{Preliminaries} \label{sec:prelims}
{
This section has definitions and preliminary results.
Our notation and terms are consistent with \cite{Diestel},
and readers are referred to that text for further information.

Let $G=(V,E)$ be a (loop-free) multi-graph with non-negative costs $c\in\Rp^{E}$ on the edges.
We take $G$ to be the input graph, and we use $n$ to denote $|V(G)|$.
For a set of edges $F\subseteq E(G)$, $c(F):=\sum_{e\in F}c(e)$,
and for a subgraph $G'$ of $G$, $c(G'):=\sum_{e\in E(G')}c(e)$.

For a positive integer $k$, we use $[k]$ to denote the set $\{1,\dots,k\}$.

For a graph $H$ and a set of nodes $S \subseteq V(H)$,
$\Gamma_H(S):=\{w\in{V(H) \setminus S}\,:\,v\in{S},vw\in{E(H)}\}$, thus,
$\Gamma_H(S)$ denotes the set of neighbours of $S$.

% For any subgraph $K$ of a graph $H$ with $V(K)\subsetneq{V(H)}$,
% an \textit{attachment} of $K$ is a node of $K$ that has a neighbour in $V(H) \setminus V(K)$.

For a graph $H$ and a set of nodes $S\subseteq V(H)$,
$\delta_H(S)$ denotes the set of edges that have one end~node in
$S$ and one end~node in $V(H) \setminus S$;
moreover,
$H[S]$ denotes the subgraph of $H$ induced by $S$, and
$H-S$  denotes the subgraph of $H$ induced by $V(H) \setminus S$.
For a graph $H$ and a set of edges $F\subseteq E(H)$,
$H-F$ denotes the graph $(V(H),~E(H) \setminus F)$.
We may use relaxed notation for singleton sets, e.g.,
we may use $\delta_H(v)$ instead of $\delta_H(\{v\})$, and
we may use $H-v$ instead of $H-\{v\}$, etc.

We may not distinguish between a subgraph and its node~set;
for example, given a graph $H$ and a set $S$ of its nodes, we use
$E(S)$ to denote the edge~set of the subgraph of $H$ induced by $S$.

% We use the standard notion of contraction of an edge, see \cite[p.25]{Schrijver}:
% Given a multi-graph $H$ and an edge $e=vw$,
% the contraction of $e$ results in the multi-graph $H/(vw)$ obtained from $H$
% by deleting $e$ and its parallel copies and identifying the nodes $v$ and $w$.
% (Thus every edge of $H$ except for $vw$ and its parallel copies
% is present in $H/(vw)$; we disallow loops in $H/(vw)$.)
% For a set of nodes $S$ of $H$ that induces a connected subgraph,
% we may use the abbreviated notation $H/S$ to denote $H/E(S)$.

%%%%%
\subsection{2EC, 2NC and related notions}

A multi-graph $H$ is called $k$-edge connected if $|V(H)|\ge2$ and for
every $F\subseteq E(H)$ of size $<k$, $H-F$ is connected.
Thus, $H$ is 2-edge connected if it has $\ge2$ nodes and the deletion
of any one edge results in a connected graph.
A multi-graph $H$ is called $k$-node connected if $|V(H)|>k$ and for
every $S\subseteq V(H)$ of size $<k$, $H-S$ is connected.
We use the abbreviations \textit{2EC} for ``2-edge connected," and
\textit{2NC} for ``2-node connected."

For any instance $H$, we use $\opt(H)$ to denote the minimum cost of a feasible subgraph
(i.e., a subgraph that satisfies the requirements of the problem).
When there is no danger of ambiguity, we use $\opt$ rather than $\opt(H)$.

By a \textit{bridge} we mean
an edge of a connected (sub)graph whose removal results in two
connected~components, and by a \textit{cut-node} we mean a node of a
connected (sub)graph whose deletion results in
two or more connected~components.
% from [Diestel]
A maximal connected subgraph without a cut-node is called a \textit{block}.
Thus, every block of a given graph $G$ is either a maximal 2NC~subgraph,
or a bridge (and its incident nodes), or an isolated node.
For any node $v$ of $G$, let $\Gamblocks{G}(v)$ denote the set of
2NC~blocks of $G$ that contain $v$.

%%%%%
\subsection{Ear decompositions}
{
An \textit{ear decomposition} of a graph 
is a partition of the edge set into paths or cycles,
$P_0,P_1,\dots,P_{\ell}$, such that
$P_0$ is the trivial path with one node,
and each $P_i$ ($1\leq i\leq {\ell}$) is either
(1)~a path that has both end~nodes in
$V_{i-1} = V(P_0) \cup V(P_1) \cup \ldots \cup V(P_{i-1})$
but has no internal nodes in $V_{i-1}$, or
(2)~a cycle that has exactly one node in $V_{i-1}$.
For an ear $P_i$, let $\intn(P_i)$ denote the set of nodes $V(P_i)\setminus{V_{i-1}}$.
Each of $P_1,\ldots,P_{\ell}$ is called an \textit{ear};
note that $P_0$ is not regarded as an ear.
We call $P_i, i\in\{1,\dots,{\ell}\},$ an \textit{open ear} if it is a path,
and we call it a \textit{closed ear} if it is a cycle.
An {\it open} ear decomposition $P_0,P_1,\ldots,P_{\ell}$
is one such that all the ears $P_2,\ldots,P_{\ell}$ are open.
(The ear $P_1$ is always closed.)

\begin{proposition}[Whitney \cite{W32}]
\label{propo:whitney}
\begin{itemize}
\item[(i)]
A graph is 2EC $\iff$ it has an ear decomposition.
\item[(ii)]
A graph is 2NC $\iff$ it has an open ear decomposition.
\end{itemize}
\end{proposition}
}

%%%%%
\subsection{Algorithms for basic computations}
There are well-known polynomial-time algorithms for implementing
all of the basic computations in this paper, see \cite{Schrijver}.
We state this explicitly in all relevant results,
but we do not elaborate on this elsewhere.

}

%%%%%%%%%%%%%%%%%%%%
\section{{\fptas} for $\kcycle$} \label{sec:fptas-steiner-cycle}
{
Bj\"orklund, Husfeldt, and Taslaman \cite{BHT12} presented a
randomized algorithm for finding a min-cost simple cycle
that contains a given set of terminals $\term$
of an unweighted, undirected graph $G=(V,E)$ with a running time of $2^k n^{O(1)}$,
where $k=|\term|$ and $n=|V|$.
In other words, they present a randomized {\fpt}-algorithm for the
unweighted $\kcycle$ problem.

\begin{theorem} \label{thm:BHTunweighted}
Consider a graph $G = (V,E)$ and a set of terminals $T \subseteq V$ of size $k$.
Let $\eta > 0$ be a parameter.
A minimum-size $\kcycle$ can be found, if one exists, by a randomized
algorithm in time {$2^k n^{O(1)} \log \frac{1}{\eta}$} with probability at least $1 - \eta$.
\end{theorem}

We present a simple (randomized) {\fptas} for the weighted $\kcycle$
problem, based on the algorithm of \cite{BHT12}.

\begin{proposition} \label{propos:BHTweighted}
Consider a graph $G = (V,E)$ with nonnegative costs $c\in\Rp^E$ on
the edges, and a set of $k$ terminals $T \subseteq V$.
Let $\ve, \eta > 0$ be some parameters.
There is a randomized algorithm that finds a $(1+\ve)$-approximate
$\kcycle$, if one exists, with probability at least $1 - \eta$.
The running time of the algorithm is {$O\Bigl(2^k \cdot n^{O(1)}
\cdot \bigl(\frac{1}{\ve}\bigr)^{O(1)} \cdot \log \frac{1}{\eta}\Bigr)$}.
\end{proposition}

\begin{proof}
Let $E = \{e_1,e_2,\dots,e_m\}$ where $c_{e_1} \leq c_{e_2} \leq \dots \leq c_{e_m}$.
Let {$\eta' := \eta/2$}. %$\epsilon' := \ve/2$.
Let $j \in [m]$ denote the smallest index such that the graph
$(V,\{e_1,\dots,e_j\})$ contains a $\kcycle$. Note that if $G$ does
not have a $\kcycle$, then the weighted-version of the problem is
trivially infeasible.
Using at most $m$ applications of Theorem~\ref{thm:BHTunweighted}
with the $\eta$-parameter set to $\eta'$, we can find the index $j$
with probability at least $1 - \eta/2$.
Suppose that we have the correct index $j$.
Let $\beta := c(e_j)$.
Let $Q^*$ denote an optimal $\kcycle$ in $G$, and $\opt := c(Q^*)$ denote the optimal cost.
By the definition of $j$, $\beta \leq \opt \leq n \beta$.
In particular, every edge in $Q^*$ has cost at most $n \beta$.
We now describe our randomized algorithm for obtaining a $\kcycle$
with cost at most $(1+\ve) \opt$.
First, we discard all edges $e$ of $G$ with cost $c_e > n\,\beta$.
Let $\mu := \ve \beta/n$; this is our ``scaling parameter".
For each edge $e$, define $\tc_e := \mu \cdot \max(1,\lceil c_e/\mu \rceil)$.
Note that $\tc_e = \mu$ if $c_e = 0$. 
\big(Observe that this rounding introduces errors, but the total error incurred
on any cycle is $\leq n\,\mu \leq$ {$\epsilon \beta \leq \epsilon\opt$}.\big)
Consider the graph $\tG = (\tV,\tE)$ obtained from $G$ by replacing
each edge $e$ by a path of $\tc_e/\mu$ edges (of unit cost).
Note that $|\tV| \leq |V| + |E| \cdot (n \beta) / \mu = O(m n^2/\ve)$.
Using a single application of Theorem~\ref{thm:BHTunweighted}, we
can obtain a minimum-size $\kcycle$ $\tQ \subseteq \tE$ with
probability at least $1 - \eta/2$ in $O\Bigl(2^k \cdot
\bigl(\frac{n^2m}{\ve}\bigr)^{O(1)} \cdot \log \frac{1}{\eta}\Bigr)$ time.
Let $Q$ denote the $\kcycle$ in $G$ corresponding to $\tQ$. 
By our choice of $\tc$, we have $c(Q) \leq \tc(Q) \leq \mu \cdot |\tQ|$.
Since the optimal $\kcycle$ $Q^*$ consists of at most $n$ edges
each with cost at most $n \beta$, the (unweighted) $\kcycle$ $\tQ^*$
in $\tG$ corresponding to $Q^*$ satisfies $\mu |\tQ^*| \leq \tc(Q^*)
\leq c(Q^*) + n \mu \leq \opt(1 + \ve)$.
By the above discussion, we can obtain a $\kcycle$ $Q$ satisfying
$c(Q) \leq \mu |\tQ| \leq \mu |\tQ^*| \leq (1+\ve) \opt$ with
probability at least $1 - \eta$.
Clearly, the overall running time is
$O\Bigl(2^k \cdot n^{O(1)} \cdot \bigl(\frac{1}{\ve}\bigr)^{O(1)}
\cdot \log \frac{1}{\eta}\Bigr)$.
\end{proof}
}
%%%%%%%%%%%%%%%%%%%%
\section{{\fptas} for $\ktwoncs$} \label{sec:fptas-k-steiner-2NCS}
{
Our (randomized) {\fptas} for the $\ktwoncs$ problem is based on a lemma
that describes the structure of a feasible subgraph,
and it repeatedly applies the algorithm of \cite{BHT12} for $\scycle$.
First, we present a randomized polynomial-time algorithm
for finding an optimal subgraph for the special case of unweighted $\ktwoncs$;
then, using the method from Section~\ref{sec:fptas-steiner-cycle}
we extend our algorithm to a (randomized) {\fptas} for weighted $\ktwoncs$.

In this section, we denote an instance of the $\ktwoncs$ problem by
$(G=(V,E),\,{c\in\Rp^{E}},\,\term\subseteq{V})$;
$G$ is the input graph with non-negative edge costs $c$, and
$\term$ is the set of terminals, $|\term|\geq3$
(we skip the easy case of $|\term|=2$).
We assume (w.l.o.g.) that $G$ is a feasible subgraph,
that is, all terminals are contained in one block of $G$.

For any graph $H$, let $D_3(H)$ denote the set of nodes that have degree $\geq3$ in $H$.

\begin{lemma} \label{lem:2ncs}
Let $H=(V',E')$ be an (edge) minimal 2NC~subgraph that contains $\term$.
Then $H$ has an open ear decomposition
$P_0,P_1,\dots,P_{\ell}$ such that
\\
(i)~each of the ears $P_i~(i\in[\ell])$ contains a terminal as an internal node
(i.e., $\intn(P_i)\cap\term\not=\emptyset$), and $P_0$ contains a terminal,
\\
(ii)~$|D_3(H)| \leq 2(|\term|-2)$.
\end{lemma}
\begin{proof}
(i)~Pick any terminal to be $P_0$.
Suppose we have constructed open ears $P_1,\dots,P_{i-1}$
and that each $\intn(P_j) (j\in[i-1])$ contains a terminal.
Let $F=\cup_{j=1}^{i-1} E(P_j)$.
Let $t$ be a terminal in $\term\setminus{V(F)}$
(we have the required ear decomposition, if $\term \subseteq V(F)$).
Suppose $i=1$;
then, $G$ has two openly disjoint paths between $t$ and $P_0$,
and we take $P_1$ to be the edge-set of these two paths.
Suppose $i\ge2$;
then, $G$ has a two-fan $P$ between $t$ and $V(F)$
(i.e., $P$ is the union of two paths between $t$ and $V(F)$
that have only the node $t$ in common);
we take $P_i$ to be $P$.
\\
(ii) Clearly, $\ell \leq |\term| - 1$ for the ear decomposition of~(i),
and each of the ears $P_2,\dots,P_{\ell}$
contributes at most $2$ (new) nodes to $D_3(H)$.
\end{proof}

The next lemma states an extension of Proposition~\ref{propo:whitney}.

\begin{lemma} \label{lem:2nc-augment}
Let $G=(V,E)$ be a graph, and let $H=(V',E')$ be a 2NC~subgraph of $G$.
Let $P$ be a path of $G$ that has both end~nodes in $V'$.
Then, $H \cup P = (V' \cup V(P), E' \cup E(P))$ is a 2NC~subgraph of $G$.
\end{lemma}

Each set $S\subseteq{V}$ of size $\leq2|\term|-4$
is a candidate for $D_3(H)$ for a 2NC~subgraph $H$ that contains $\term$, and
we call $\term\cup{S}$ the set of \textit{marker} nodes.

Our algorithm has several nested loops.
The outer-most loop picks a set $S\subseteq{V}$ of size $\leq2|\term|-4$,
and then applies the following main~loop.
Each iteration of the main~loop attempts to construct a 2NC~subgraph
that contains the set of marker~nodes $\term\cup{S}$,
by iterating over all ordered partitions
$(\pterm_1,\pterm_2,\dots,\pterm_r)$ of $\term\cup{S}$ such that $|\pterm_1|\ge2$
and the number of sets in the partition, $r$, is a positive integer, $r\leq{k}={|\term|}$.

Consider one of these ordered partitions
$(\pterm_1,\pterm_2,\dots,\pterm_r)$.
We attempt to find a min-cost $\scycle$ $C_1$ that contains $\pterm_1$
using the algorithm of \cite{BHT12};
if $G$ has no $\scycle$ that contains $\pterm_1$,
then this iteration has failed, otherwise,
we take $C_1$ to be the first (closed) ear of an open ear decomposition of
our candidate 2NC~subgraph that contains $\term\cup{S}$.
Then, for $i=1,\dots,r-1$, we pick a pair of nodes
$s_i,t_i \in \pterm_1\cup\dots\cup\pterm_i$, and attempt to find
a min-cost $\spath$ $P_{i+1}$ between $s_i$ and $t_i$ that contains $\pterm_{i+1}$;
if $G$ has no such $\spath$, then this iteration has failed, otherwise,
we augment the current subgraph $H:=C_1\cup{P_2}\cup\dots\cup{P_i}$ by $P_{i+1}$.

The algorithm maintains an edge-set $\hF$;
initially, $\hF=E$, and, at termination,
$\hF$ is the edge-set of a min-cost 2NC~subgraph that contains $\term$.

Pseudo-code for the algorithm is presented below.

We use {$\A_{\bhtcycle} (G,\;\pterm_1{,\;\eta})$} to denote a call to
the $\scycle$ algorithm of \cite{BHT12} where the inputs are the graph $G$,
the terminal set $\pterm_1 \subseteq V(G)$, {and the desired probability of failure $\eta$}.
With probability {at least $1-\eta$},
this call either returns the edge-set of a minimum-size cycle of $G$
that contains all nodes of $\pterm_1$ 
or reports an error if $G$ has no such cycle.

We use {$\A_{\bhtpath} (G,\;\pterm_1,\;s,\;t{,\;\eta})$} to denote a call to the following subroutine
that attempts to find an $s,t$-path of $G$ that contains all nodes of $\pterm_1$.
First, construct an auxiliary graph $G'$ from $G$ by adding a node $u'$
and two edges $u's, u't$. Then call $\A_{\bhtcycle} (G',\;\pterm_1\cup\{u',s,t\}{,\;\eta})$;
report an error if the call returns an error, and, otherwise,
return the path obtained by deleting the node $u'$ (and its two incident edges)
from the cycle returned by the call.

%-----
\begin{figure} 
\begin{minipage}{0.9\textwidth}
\textbf{Algorithm } $\A_{2NC}(G, \,   \term, \, \eta)$ \\
({Set parameter}) {$\eta' = \frac{\eta}{k}$}; \\
({Initialize}) $\hF := E$;\\
\textsc{For } $S \subseteq V$ such that $|S|\leq 2k$:\\
\mbox{~}\quad\textsc{For } $r = 1,\ldots,{k}$:\\
\mbox{~}\quad\quad \textsc{For} ordered partitions $(\pterm_1,\ldots,\pterm_r)$ of $\term\cup S$ such that $|\pterm_1|\geq 2$:\\
\mbox{~}\quad\quad\quad \textsc{For} $i=1,\ldots,r-1$ and
	node pairs $(s_i,t_i) \in \cup_{j=1}^i \pterm_i$,
	where $s_i\neq t_i$:\\
\mbox{~}\quad\quad\quad\quad $H := \A_{\bhtcycle} (G,\pterm_1{,\;\eta'}) \quad
		\bigcup _{i=1}^{r-1} \A_{\bhtpath}(G,\pterm_{i+1},s_i,t_i{,\;\eta'})$;\\
\mbox{~}\quad\quad\quad\quad \textsc{Continue } the loop if
		any call to any subroutine reports an error; \\
\mbox{~}\quad\quad\quad\quad \textsc{If } $|E(H)|<|\hF|$, update $\hF := E(H)$;\\
\mbox{~}\quad\quad\quad \textsc{End For};\\
\mbox{~}\quad\quad \textsc{End For};\\
\mbox{~}\quad \textsc{End For};\\
\textsc{End For};\\
\textsc{Output } $\hF$;\\
\end{minipage}
\label{alg:2NC}
\end{figure}
%-----

\begin{lemma} \label{lem:2ncs-correct}
Let $H^*=(V^*,E^*)$ be an optimal subgraph for $\ktwoncs$.
Assume that each of the calls to the subroutines
(namely, $\A_{\bhtcycle},\A_{\bhtpath}$)
returns a valid subgraph whenever one exists.
Let $H=(U,\hF)$ denote the output of the above algorithm.
Then $H$ is a 2NC~subgraph, $U\supseteq{\term}$, and $|\hF|\leq |E^*|=\opt$.
\end{lemma}
\begin{proof}
By Lemma~\ref{lem:2ncs},
$H^*$ has an open ear decomposition $P_1,P_2,\ldots,P_{r^*}$ such that
each of the ears $P_i$ contains at least one terminal as an internal node;
hence, $r^* \leq k=|\term|$.
Let $S^* = D_3(H^*)$ be the set of nodes of degree $\geq3$ of $H^*$;
clearly, $|S^*| \leq 2r^* \leq 2k$.

For $i=1,\ldots, r^*$, let $T^*_i = P_i \cap (\term\cup S^*)$.
For $i=1,\ldots,r^*-1$, let $(s^*_i,t^*_i)$ denote the end~nodes of the ear $P_{i+1}$;
clearly, $(s_i^*,t_i^*) \in \cup_{j=1}^{i}(T_i^*)$. 

Now consider the loop in the algorithm where
$S = S^*$, $r=r^*$, $\pterm_i = T^*_i$ for $i=1,\ldots, r^*$ and
$(s_i,t_i) = (s^*_i,t^*_i)$ for $i=1,\ldots,r^*-1$.
Observe that the calls to the subroutines $A_{\bhtcycle}$ and $\A_{\bhtpath}$
return minimum-size subgraphs, hence,
$|\A_{\bhtcycle}(G,\pterm_1)|\leq |P_1|$ and
$|\A_{\bhtpath}(G,\pterm_{i+1},s_i,t_i)| \leq |P_{i+1}|$ for $i=1,\ldots,r^*-1$.
Since $|E^*| = \sum_{i=1}^{r^*}|P_i|$, we conclude that the
2NC~subgraph $H$ found by this iteration satisfies $|E(H)|\leq
|E^*|$. Thus, the algorithm outputs an optimal 2NC~subgraph that
contains $\term$.
\end{proof}

\begin{theorem} \label{thm:2ncs}
Consider the unweighted $\ktwoncs$ problem with $k=O(1)$.
Let $\eta > 0$ be a parameter.
With probability $\geq 1 - \eta$,
the above randomized algorithm computes an optimal subgraph in time
	{$O(\binom{n}{2k}\cdot B(3k)\cdot { \binom{3k}{2}^{k}} \cdot 2^{3k}n^{O(1)}
	\log{\frac{k}{\eta}}) =
	O\left(n^{O(k)} \cdot \log{\frac{1}{\eta}}\right)$},
where $B(i)$ denotes the $i^{th}$ ordered Bell number.
\end{theorem}
\begin{proof}
{{As seen in the proof of Lemma \ref{lem:2ncs-correct}, if the subroutines $A_{\bhtcycle}$ and $\A_{\bhtpath}$ run correctly when $S = S^*$, $r=r^*$, $\pterm_i = T^*_i$ for $i=1,\ldots, r^*$ and
$(s_i,t_i) = (s^*_i,t^*_i)$ for $i=1,\ldots,r^*-1$ corresponding to an ear decomposition of an optimal solution $H^*$, then the above algorithm outputs an optimal solution. During this loop, there are at most $r^*\leq k=|\term|$ calls to the subroutines $A_{\bhtcycle}$ and $\A_{\bhtpath}$. Hence, with probability at least $(1-\eta')^{k} \geq 1-\eta$,
%the above algorithm 
{Algorithm~$\A_{2NC}$ 
outputs an optimal solution.}
}
%By the union bound, the probability that any of the calls to the subroutines fail is at most $1-N\eta'=1-\frac{N\eta}{n^{O(k)}}$ where $N$ is the number of times that a subroutine is called. It follows from the runtime analysis below that $N < n^{O(k)}$. Suppose that none of these calls fail.} Then, by Lemma~\ref{lem:2ncs-correct}, the algorithm finds an optimal subgraph.

The running time is analyzed as follows:
the term $\binom{n}{2k}$ comes from choosing $S \subseteq V,\,|S|\leq2k$
(in the outer-most loop),
the term $B(3k)$ comes from choosing ordered partitions of $S\cup\term$,
the term ${\binom{3k}{2}^{k}}$ comes from choosing the node pairs $(s_i,t_i)$ 
for {the $r-1 (\le k)$ calls} to $\A_{\bhtpath}$,
and the term {$2^{3k}n^{{O(1)}}{\log\frac{k}{\eta}}$} comes from the running time of the
algorithm of \cite{BHT12} for the $\scycle$ problem, {with error probability {$\frac{\eta}{k}$}}.
}
\end{proof}

{
\begin{corollary} \label{coro:2ncs-weighted}
There is a randomized algorithm for the (weighted) $\ktwoncs$
problem that runs in time
\[ 
    {O\left(n^{O(k)}\cdot \left(\frac{1}{\ve}\right)^{O(k)}\cdot\log{\frac{1}{\eta}}\right)}
\]
such that, with probability at least $1-\eta$, the cost of the solution returned by the
algorithm
is at most $(1+\ve)$ times the cost of an optimal solution.
\end{corollary}
}
%%%%%%%%%%%%%%%%%%%%
\section{{\fptas} for $\kfst$ and $\ktwoecs$} \label{sec:fptas-k-fst}
{
In this section we present a randomized polynomial-time algorithm
for finding an optimal subgraph for the special case of unweighted
$\kfst$; then, using the method from Section~\ref{sec:fptas-steiner-cycle},
we extend our algorithm to a (randomized) {\fptas} for weighted $\kfst$.
We assume that $k=|\term|\geq3$ is a positive integer.
Note that the $\ktwoecs$ problem is a special case of $\kfst$ where
all edges are unsafe.

Adjiashvili, Hommelsheim, M\"{u}hlenthaler, and Schaudt \cite{AHMS20}
give a polynomial-time algorithm for finding an optimal solution to the
$\twofst$ problem; see Proposition~1 and Theorem~5 of \cite{AHMS20}.
We refer to their $\twofst$ algorithm as $\A_{\text{2-FST}}$.
We refer to (inclusion-wise) minimal feasible solutions to a $\twofst$
problem on $G$ as \textit{\ahmspaths}.

Informally speaking, our randomized polynomial-time algorithm for
$\kfst$ represents minimal feasible solutions as 2NC~blocks connected
together using \ahmspaths.
To simplify our presentation, we first modify the $\kfst$ instance
$G = (V,\safe\sqcup\unsafe,\term)$ as follows.
For each terminal $v\in\term$, we create a new node $v'$ and a new
safe edge $vv'$. Let $\term'$ denote the set of these new nodes and
let $E'$ denote the set of the new safe edges. Consider the modified
instance $G' = (V\cup\term',(\safe\cup{E'})\sqcup\unsafe,\term')$.
Observe that $(U,F)$ is a feasible solution to the original instance
if and only if $(U\cup\term',F\cup{E'})$ is a feasible solution to
the modified instance.

\begin{definition}[Block-Tree]
A block-tree of a graph $G$ is a tree $\BT{G}$ with the following properties:
\begin{enumerate}
    \item The nodes of $\BT{G}$ are in one-to-one correspondence
    with the 2NC~blocks of $G$.
    \item If two 2NC~blocks are connected by a bridge in $G$, then
    the two corresponding nodes in $\BT{G}$ are adjacent.
    \item
	For each cut-node $v$ of $G$, the subgraph of $\BT{G}$
	induced by $\Gamblocks{G}(v)$ is connected ($\Gamblocks{G}(v)$ is
	the set of 2NC~blocks of $G$ that contain $v$).
	In other words, the unique path of $\BT{G}$ between any two nodes of
	$\Gamblocks{G}(v)$ has all its internal nodes in $\Gamblocks{G}(v)$.
%
%	For any node $v$ in $G$, the set of all nodes in $\BT{G}$ corresponding
%	to 2NC~blocks that contain $v$ forms a connected subtree of $\BT{G}$.
\end{enumerate}
\end{definition}
Informally speaking, a block-tree of a graph $G$ represents how the
2NC~blocks of $G$ are connected together.  Each edge of the block-tree
either represents a bridge of $G$ or connects a pair of 2NC~blocks
of $G$ that share a common cut-node.  Let $H$ be a minimal feasible
$\kfst$ solution. Due to the modification above, we may assume that
every leaf of $\BT{H}$ corresponds to a block of $H$ that contains
exactly one terminal.
Then any path in $\BT{H}$ corresponds to
a $\ahmspath$ of $H$ that connects either (i)~two cut-nodes, or (ii)~a
cut-node and a terminal, or (iii)~two terminals.

For our algorithmic application, nodes of $\BT{H}$ of degree two
are redundant, and this motivates the notion of a ``non-redundant''
block-tree.

\begin{definition}[Condensed Block-Tree]
A condensed block-tree of a graph $G$ is a tree $\CBT{G}$ obtained
from a block-tree $\BT{G}$ with the following properties:

\begin{enumerate}
    \item The nodes of $\CBT{G}$ are nodes $b$ of $\BT{G}$ such that
	$\deg_{\BT{G}}(b) \neq 2$.
    \item Two nodes $b_1$ and $b_2$ are adjacent in $\CBT{G}$ if
    and only if every internal node in the path connecting $b_1$
    and $b_2$ in $\BT{G}$ has degree~two.
\end{enumerate}
\end{definition}
\begin{figure}[htb]
\centerline{\includegraphics[scale=0.30]{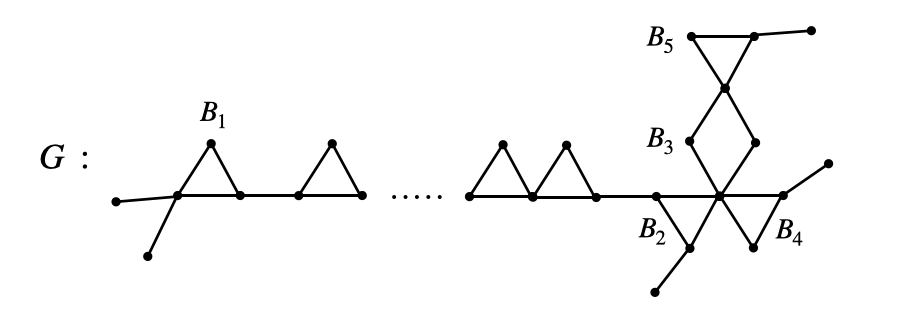}
}
\caption{The original graph $G$}
\centerline{\includegraphics[scale=0.30]{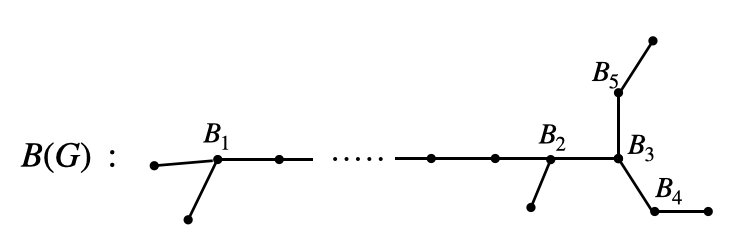} \includegraphics[scale=0.30]{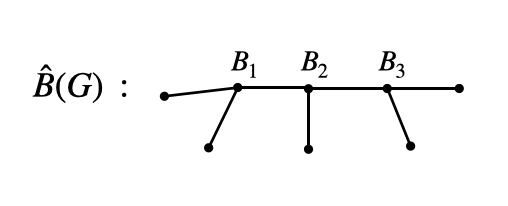}}
\caption{A block tree $\BT{G}$ and the corresponding condensed block tree $\CBT{G}$}
\end{figure}
For any minimal feasible solution $H$ of $\kfst$ and a condensed
block-tree $\CBT{H}$, we refer to the 2NC~blocks of $H$ that
correspond to internal nodes of $\CBT{H}$ as \textit{high-degree~blocks}.
The leaves of $\CBT{H}$ correspond to the terminals.
Edges of $\CBT{H}$ correspond to $\ahmspaths$ in $H$ that connect either
(i)~two high-degree~blocks, or
(ii)~a high-degree~block and a terminal, or
(iii)~two terminals.
The end-points of these $\ahmspaths$ are either cut-nodes of $H$ or terminals.
Note that some of these $\ahmspaths$ could be trivial~paths
corresponding to cut-nodes that are common to two high-degree~blocks.
We now state some useful lemmas that follow from the handshaking
lemma applied to $\CBT{H}$.

\begin{lemma}\label{lem:BoundHighDegreeBlocks}
The number of internal nodes (i.e., non-leaf nodes) of $\CBT{H}$
is at most {$k-2$} where $k$ is the number of terminals.
\end{lemma}

\begin{lemma}\label{lem:BoundCutNodes}
The total number of cut-nodes (with repetitions) in high-degree~blocks
of $H$ is at most {$3k-6$} where $k$ is the number of terminals.
\end{lemma}

Now, we describe our algorithm for unweighted $\kfst$.
We guess (via enumeration) the high-degree~blocks of an optimal
solution $\optsoln$ corresponding to some condensed block-tree $\CBT{\optsoln}$.
The guess would include the number of high-degree~blocks and the
cut-nodes in each of these high-degree~blocks. This is done by
picking {$3k-6$} nodes of $V$ with replacement and then picking a partition $\cp$
of these {$3k-6$} nodes into at most {$k-2$} sets.
Let {$r\leq k-2$} be the number of sets in the partition $\cp$.
Thus, $\cp=(X_1,X_2,\dots,X_r)$.
For each set $X_i$, we use algorithm $\A_{2NC}$ to find $B_i$, a
minimum size 2NC~subgraph of $G$ containing the specified cut-nodes
in $X_i$, possibly, with some additional Steiner nodes.  Finally,
we construct a tree that connects these 2NC~subgraphs and terminals
via $\ahmspaths$ using the following subroutine.

First, for every pair of nodes $(u,v)\in V'\times V'$, we use
algorithm $\A_{\text{2-FST}}$ to find $\mG{uv}$, the minimum size
$\ahmspath$ connecting $u$ and $v$ in $G'$. We then construct a
complete graph $K(X_1,\ldots,X_r)$ with $r+k$ nodes that  has one
node for each set $X_i$ and one node corresponding to each
terminal $\{t\}$. The cost of an edge between two nodes of $K$
corresponding to node sets $V_1$ and $V_2$ is given by
$\min\{|E(\mG{uv})|: u\in V_1, v\in V_2\}$. Note that if there is
no $\ahmspath$ connecting a node in $V_1$ to a node in $V_2$,
then we fix the cost of the edge to be infinity. Thus an edge $\Bar{e}$ of $K$
corresponds to a subgraph $\mG{\Bar{e}}$ in $G'$ which is the
minimum size $\ahmspath$ whose end points are in $V_1$ and $V_2$
respectively. We then find a minimum spanning tree $\Bar{T}$ in $K$.
Note that if $\Bar{T}$ has infinite cost, then we output an error.
Else, we output the subgraph of $G'$ defined by $\mG{}(X_1,\ldots,X_r)
:= \cup_{\Bar{e}\in \Bar{T}} \mG{\Bar{e}}$.

\begin{center} \label{alg:kFST}
\begin{minipage}{\textwidth}
\textbf{Algorithm } $\A_{\text{k-FST}}( G',\; \term'{,\;\eta})$\\
{({Set parameter}) {$\eta' = \frac{\eta}{k}$}; \\}
({Initialize}) $H := G'$;\\
\textsc{For } {$S=(v_1,\ldots,v_{3k-6}) \in V^{3k-6}$}:\\
\mbox{~}\quad\textsc{For } {$r = 1,\ldots,k-2$}:\\
\mbox{~}\quad\quad \textsc{For} partitions $(X_1,\ldots,X_r)$ of $S$:\\
\mbox{~}\quad\quad\quad $\hH := \mG{}(X_1,\ldots,X_r) \cup_{i=1}^r \A_{2NC}(G,X_i{,\;\eta'})$;\\
\mbox{~}\quad\quad\quad \textsc{Continue } the loop if
		any call to any subroutine reports an error; \\
\mbox{~}\quad\quad\quad \textsc{If } $|E(\hH)|<|E(H)|$, update $H := \hH$;\\
\mbox{~}\quad\quad \textsc{End For};\\
\mbox{~}\quad \textsc{End For};\\
\mbox{~}\textsc{End For};\\
\textsc{Output } $H$;\\
\end{minipage}
\end{center}

\begin{lemma}\label{lem:kFSTcorrect}
Let $H^*=(V^*,E^*)$ be an optimal subgraph for $\kfst$. 
Assume that each of the calls to the subroutine {$\A_{2NC}(G,X_i,\eta')$}
returns a valid subgraph {${2NC}(G,X_i)$} whenever one exists.
Let $H = (U,F)$ denote the output of the above algorithm. Then,
$H$ is a feasible $\kfst$ solution and $|F|\leq |E^*| = \opt$.
\end{lemma}

\begin{proof}
We argue that the subgraph $\hH$ in any iteration of the algorithm is a
feasible $\kfst$ solution. This holds because the algorithm finds
2NC~subgraphs {${2NC}(G,X_i)$} and then connects them
to one another and to the terminals using the $\ahmspaths$ in
$\mG{}(X_1,\ldots,X_r)$. Thus, $\term\subseteq V(\hH)$ and any unsafe edge
$e\in E(\hH)$ either lies in a 2NC~subgraph of $\hH$ or a
$\ahmspath$ in $\hH$, hence, $\hH-e$ is connected.

Now consider a condensed block-tree $\CBT{H^*}$. Let
$B^*_1,\ldots,B^*_{r^*}$ be the high-degree~blocks of $H^*$ and let
$X^*_i$ be the set of cut-nodes in $B^*_i$. By
Lemma~\ref{lem:BoundCutNodes}, the total number of cut-nodes in all the
high-degree~blocks $B^*_i$ is at most {$3k-6$}. We may assume that
it is exactly {$3k-6$} by duplicating a cut-node $v \in {X^*_1}$
multiple times. Consider the iteration of the algorithm where $r=r^*$
and $X_i = X^*_i$ for $i=1,\ldots, r^*$. Then,
\begin{align*}
    {|E({2NC}(G,X_i))|} \leq |E(B^*_i)| \quad\quad \forall i=1,\ldots,r.
\end{align*}
Recall that the nodes of $\CBT{H^*}$ correspond to the high-degree~blocks
$B^*_i$ (and hence to the node sets $X^*_i$) or to the terminals
$\{t\}$. Also an edge $\Bar{e}$ of $\CBT{H^*}$ between nodes
corresponding to node sets $V_1$ and $V_2$ represents a $\ahmspath$ $H^*_{\Bar{e}}$
in $H^*$ whose end-points lie in $V_1$ and $V_2$ respectively. Hence
$\CBT{H^*}$ may be viewed as a subgraph of $K(X_1,\ldots,X_r)$.
Furthermore, since any two nodes in $H^*$ have a $\ahmspath$
between them, $\CBT{H^*}$ must be connected. Finally, by construction
of $K$, $|E(H^*_{\Bar{e}})| \geq c(\Bar{e})$ where $c(\Bar{e})$ is
the cost of the edge $\Bar{e}$ in $K$. This implies that the cost
of the minimum spanning tree in $K$ is at most $\sum_{\Bar{e}\in
\CBT{H^*}}|E(H^*_{\Bar{e}})|$. Hence,
\begin{align*}
{|E(\mG{}(X_1,\ldots,X_r))|} \leq \sum_{\Bar{e}\in \CBT{H^*}}|E(H^*_{\Bar{e}})|.
\end{align*}
Combining the two inequalities above we obtain 
\begin{align*}
    {|E(\hH)|} &= |E({\mG{}(X_1,\ldots,X_r)} \cup_{i=1}^r {{2NC}(G,X_i))}| \\
    &\leq {|E(\mG{}(X_1,\ldots,X_r))|} + \sum_{i=1}^r {|E({2NC}(G,X_i))|} \\
    &\leq \sum_{\Bar{e}\in \CBT{H^*}}|E(H^*_{\Bar{e}})| + \sum_{i=1}^r |E(B^*_i)| \\
    &= |E^*|
\end{align*}
The last equation holds because $E^*$ partitions into
the edge-sets of the high-degree~blocks $B^*_i$
and the edge-sets of the $\ahmspaths$ $H^*_{\Bar{e}}$.
This completes the proof of the lemma.
\end{proof}

\begin{theorem} \label{thm:k-fst-unweighted}
Consider the unweighted $\kfst$ problem with $k=O(1)$.
{
Let $\eta > 0$ be a parameter.
Let $\alpha$ denote the running time of $\A_{\text{2-FST}}$ and let
$\beta$ denote the running time of $\A_{\text{2NC}}(\cdot, \cdot, \eta')$ with parameter {$\eta' = \eta/k$}.
}
With probability $\geq 1 - \eta$,
the above randomized algorithm computes an optimal subgraph in time
\[
    {O(n^2\alpha \cdot {n^{3k-6}}2^k n^2\beta \cdot k^2\log{k})
    = O\left(\alpha \cdot \beta \cdot n^{3k}\right)}.
\]
\end{theorem}

{
Using the running time for $\A_{\text{2NC}}$ given above and the algorithm of
\cite{AHMS20} for $\A_{\text{2-FST}}$, which has a $n^{O(1)}$ running time, the
running time of the above algorithm for $\kfst$ is 
{$O\left(n^{O(k)} \cdot\log{\frac{1}{\eta}}\right)$}. This leads
to the following result for the weighted $\kfst$ problem.
\begin{corollary} \label{coro:k-fst-weighted}
There is a randomized algorithm for the (weighted) $\kfst$ problem that runs in time
\[ 
    {O\left(n^{O(k)} \cdot \left(\frac{1}{\ve}\right)^{O(k)} \cdot
	\log{\frac{1}{\eta}} \right)}
\]
such that, with probability at least $1 - \eta$, the solution returned by the
algorithm has cost at most $(1+\ve)$ times the cost of an optimal solution. 
\end{corollary}
}
}

\bigskip
\bigskip
\bigskip

\noindent
\textbf{Acknowledgements}. We are grateful to several colleagues for insightful comments at the start of this project.
%%%%%%%%%%%%%%%%%%%%
%%%%%%%%%%%%%%%%%%%%
\vfill
\newpage
%% Please use bibtex, 
% \bibliographystyle{plain}
\bibliographystyle{plainurl}% the mandatory bibstyle for LIPIcs
%1aug22 \bibliography{bcgi-arxiv-ref}
\bibliography{k-fst-v2}
\end{document}